\documentstyle[prb,aps,floats,epsfig]{revtex}
\newcommand{\ee}{\end{equation}}
\begin{document}
\draft
\par
\begingroup

{\large\bf\centering\ignorespaces
Gauge invariant grid discretization of Schr\"odinger equation 
\vskip2.5pt}
{\dimen0=-\prevdepth \advance\dimen0 by23pt
\nointerlineskip \rm\centering
\vrule height\dimen0 width0pt\relax\ignorespaces
Michele Governale and Carlo Ungarelli\par }

{\small\it\centering\ignorespaces
Dipartimento di Ingegneria dell'Informazione\\
via Diotisalvi 2, I-56126 Pisa, Italy. \par}

\par
\bgroup
\leftskip=0.10753\textwidth \rightskip\leftskip
\dimen0=-\prevdepth \advance\dimen0 by17.5pt \nointerlineskip
\small\vrule width 0pt height\dimen0 \relax

Using the Wilson formulation of lattice gauge theories, a gauge invariant 
grid discretization of a one-particle Hamiltonian in the presence of an 
external electromagnetic field  is proposed. 
This Hamiltonian is compared both with that obtained by a 
straightforward discretization of the continuous Hamiltonian by means 
of balanced difference methods, and with a tight-binding Hamiltonian.  
The proposed Hamiltonian and the balanced difference one 
are used to compute the energy spectrum of a charged particle in a
two-dimensional parabolic potential in the presence of a
perpendicular, constant magnetic field. 
With this example we point out  how a ``na\"{\i}ve''
discretization gives rise to an explicit breaking of the gauge invariance
and to large errors in the computed eigenvalues and
corresponding probability densities; in particular, the error on the 
eigenfunctions may lead to very poor estimates of the mean values of some 
relevant physical quantities on the corresponding states. 
On the contrary, the proposed discretized Hamiltonian allows a reliable 
computation of both the energy spectrum and the probability densities. 

\par\egroup

\thispagestyle{plain}
\pacs{\tt PACS number(s): 73.20Dx, 72.15Rn}
\endgroup
\section{Introduction}
In the last few years, there has been increasing activity
concerning numerical simulations of low energy quantum systems and in
this context particular attention has been paid to mesoscopic systems
such as low-dimensional semiconductor nanostructures. 
Of special interest are ``quasi zero'' dimensional systems; in this
case, the electrons are confined in a small region of space --as in an 
atom-- and hence they exhibit a quantized energy spectrum. In
particular, urged by many experimental results~\cite{Hans}, theoretical 
investigations of the influence of a magnetic field on the spectrum of
such systems of confined electrons  have been carried out~\cite{kum}. 

The problem of computing the energy spectrum of such systems has been 
addressed in many ways~\cite{Bryant,JoKi,Stopa2,Mac95,Stopa1,Mac97}. 
In most of the approaches used the resulting Schr\"odinger 
equation is solved numerically, introducing  a discretization grid and  
using finite difference methods (see
e.g.~\cite{Mac97}). In order to study the 
effect due to the presence of a magnetic field, the common procedure
is to introduce a vector potential (``gauge field'') 
and eventually construct an appropriate discretization of the
corresponding Hamiltonian. On the other hand, it is well known from
lattice quantum field theory that the presence of a gauge field  
may lead to the  
breaking of the  gauge invariance if the Hamiltonian  is naively
discretized on a grid. In particular the eigenvalues of the Hamiltonian 
and the probability density associated with a given eigenfunction (i.e.
its square modulus) 
could depend on the gauge choice.

The aim of this paper is to present a method of constructing 
a grid-discretized one-particle Hamiltonian in
the presence of an external electromagnetic field, that leads to a manifest gauge 
invariance of the mean values of the observables on its eigenstates. 
In order to do so we apply the Wilson formulation of lattice 
gauge theories~\cite{Wilson} (for a review on this topic see also~\cite{lrev}).

The paper is organized as follows: in
Sec.~\ref{sec1}, we construct explicitly a lattice discretization of
the one particle Hamiltonian in the presence of an external electromagnetic
field using the above mentioned formulation and we compare it with
that obtained by introducing directly the vector potential and with 
a tight-binding one. In particular, we write the proposed Hamiltonian 
in the tight-binding formalism and we show that it leads to a 
simpler expression for the hopping potential. 
 In Sec.~\ref{sec2} we use 
the ``naive'' discretized Hamiltonian and the gauge-invariant one 
to compute the energy spectrum of a charged  particle in a
two-dimensional parabolic
potential in the presence of a perpendicular, constant magnetic field ,
showing explicitly that the Wilson formulation preserves 
the gauge invariance. Finally, Sec.~\ref{sec3} contains some 
concluding remarks. 

\section{Gauge invariant lattice Hamiltonian}
\label{sec1}
In this section we apply the Wilson formulation to find a Hamiltonian
$H_{{\rm LATT}}$ defined on a lattice of discretization points that in
the continuum limit (i.e. lattice spacing going to zero) 
tends to the following generic Hamiltonian
operator for a particle with charge $q$ in the presence of a 
magnetic field $\vec B(\vec r\,) =\vec\nabla\wedge\vec A(\vec r\,) $ and of an electrostatic
potential $V(\vec r\,)$:
\begin{equation}
\label {Hcont}
H_{{\rm CONT}}={\hbar^2\over 2 m} \, \left[-\nabla ^2  +i g\left\{ \vec A(\vec r\,) ,\vec{\nabla}  \right\} 
+g^2\,\left|\vec A(\vec r\,) \right|^2\right]+V(\vec r\,) \, ,
\end{equation}
where $g=q/\hbar$,   
and the symbol 
$\{ \vec A(\vec r\,) ,\vec{\nabla}  \}$ stands 
for the sum of the anticommutators for the components of $\vec A$ 
and $\vec{\nabla}$. Throughout this paper, for simplicity we neglect
the spin. 

The essential requirement we impose is that the lattice Hamiltonian 
$H_{{\rm LATT}}$ maintains all the properties of the continuum one for a 
gauge transformation of the vector potential. 

It is well known from elementary quantum mechanics that when the
vector potential transforms as 
\begin{equation}
\vec{A} \rightarrow \vec{A} + \vec{\nabla}  \,\Lambda(\vec r\,) 
\label{gtr}
\end{equation}
a generic wave function $\Psi(\vec r\,) $, corresponding to a physical state of the system, transforms as 
\begin{equation}
\Psi(\vec r\,)  \rightarrow G(\vec r\,)  \Psi(\vec r\,)  \,\,,\quad\mbox{with}\quad G(\vec r\,) =\exp\{ig\Lambda(\vec r\,) \}\,,\label{trasf}
\end{equation}
leaving unchanged the mean values of the observables on $\Psi$. 
In the following, we shall refer to the above
mentioned property as gauge invariance. More formally, under the
gauge transformation~(\ref{gtr}) the Hamiltonian 
$H_{{\rm CONT}}$ behaves as 
\begin{equation}
H_{{\rm CONT}} \rightarrow G(\vec r\,) \,H_{{\rm CONT}}\,G^{\dagger}(\vec r\,) \,,
\label{Htr}
\end{equation}
and, as a wave function $\Psi$ transforms according to~(\ref{trasf}), 
the energy functional 
\begin{equation}
{\cal E}_c[\Psi]=\int d\vec{r}\,\Psi^{\dagger}(\vec r\,) \,H_{{\rm CONT}}\,\Psi(\vec r\,) 
\label{enc}
\end{equation}
is left unchanged.

We now introduce a uniform discretization grid with lattice constant
$a$, such that a generic point on it is identified with a vector
$x=(la,ma,na)$; we shall also indicate a generic direction
in the discretized space with the Greek letter $\mu$ . Given
a vector potential with components  $A_{\mu}(x)$, following~\cite{Wilson} we
define on the lattice the operators $U_{\mu}(x)$ as follows: 
\begin{equation}
\label{Umu}
U_{\mu}(x)=\exp\{iga\,A_{\mu}(x)\}\,.
\end{equation}

Let us consider the discrete gauge transformation corresponding to
~(\ref{gtr}), i.e.
\begin{equation}
A_{\mu} \rightarrow A_{\mu}+\frac{1}{a}\left[\Lambda(x+\mu)-\Lambda(x)\right]\,,  
\label{lgt1}
\end{equation} 
where $x+\mu$ denotes the next neighbor of $x$ in the $\mu$ direction. 
It is straightforward to show that under gauge transformations~(\ref{lgt1})  
$U_{\mu}$ behaves as 
\begin{equation}
U_{\mu}(x) \rightarrow G^{\dagger}(x)\,U_{\mu}(x)\,G(x+\mu)
\,\,,\quad\mbox{with}\quad G(x)=\exp\{ig\Lambda(x)\}\,.
\label{lgt3}
\end{equation} 
In the following, we represent the discretized Hamiltonian 
operator $\hat H_{{\rm LATT}}$ with its matrix elements between the functions 
$\chi (x-x_i)=1/\sqrt{a^3}
\delta_{x,x_i}$, where $\delta_{x,x_i}$ is equal to zero anywhere on 
the grid except at 
the lattice point $x_i$ where it equals $1$, and $1/\sqrt{a^3}$ is a 
normalization factor.
Hence we define $H_{{\rm LATT}}(x,y)$ as:
\begin{equation}
H_{{\rm LATT}}(x,y)=a^3\,\sum_{\bar x}\,\chi^\dagger (\bar x-x)\hat H_{{\rm LATT}}
\chi (\bar x-y)=\sum_{\bar x}\delta_{\bar x,x}\hat H_{{\rm LATT}}\delta_{\bar x,y}.
\end{equation}
With the definition above we have:
\begin{equation}
\hat H_{{\rm LATT}}\Psi(x)=\sum_{y}H_{{\rm LATT}}(x,y)\Psi(y).
\end{equation}
We now want to construct a lattice Hamiltonian which transforms as 
\begin{equation}
H_{{\rm LATT}}(x,y) \rightarrow G(x)H_{{\rm LATT}}(x,y)G^{\dagger}(y)\,.
\label {htr}
\end{equation}
As in the continuum case, this implies that a wave function $\Psi(\vec r\,) $ 
transforms according to the discrete analog of~(\ref{trasf}):
\begin{equation}
\Psi(x) \rightarrow G(x) \Psi(x) \,.
\label{lgt2}
\end{equation}
We find that the following Hamiltonian
\begin{equation}
H_{{\rm WH}}(x,y)={\hbar^2\over 2 m}\,\sum_{\mu}\,\frac{1}{a^2}
\left[2\delta_{x,y}-U_{\mu}(y)\,\delta_{x-\mu,y}-
U_{\mu}^{\dagger}(x)\,\delta_{x+\mu,y}\right]+V(x)\delta_{x,y}
\label{Hlat}
\end{equation}
has the correct continuum limit~(\ref{Hcont}) and it transforms as
required by Eq.~(\ref{htr}). Moreover the Hamiltonian~(\ref{Hlat})
gives rise, as in the continuum case, to a gauge invariant energy
functional
\begin{equation}
{\cal E}_{{\rm LATT}}[\Psi]=a^3\,
\sum_{x,y}\Psi^{\dagger}(x)\,H_{{\rm WH}}(x,y)\Psi(y)\,. 
	\label{end}
\end{equation} 
In order to show the advantages of the above formulation in respect to 
a straightforward discretization of~(\ref{Hcont}) we 
now consider the discrete Hamiltonian  
obtained from~(\ref{Hcont}) applying balanced difference 
methods, which reads:
\begin{eqnarray}
&&H_{{\rm BDH}}(x,y)={\hbar^2\over 2 m}\sum_{\mu}
\frac{1}{a^2}\,\biggl\{\left[2+\frac{iga}{2}\,(A_\mu(x+\mu)
-A_\mu(x-\mu))+g^2a^2\,A_\mu^2(x)\right]
\delta_{x,y}\nonumber \\
&&-\left(1+iga\,A_\mu(x)\right)
\delta_{x-\mu,y}
-\left(1-iga\,A_\mu(x)\right)\delta_{x+\mu,y}
\biggr\}+V(x)\delta_{x,y}.
\label{hwr}
\end{eqnarray}
One of the inconveniences of the Hamiltonian~(\ref{hwr}) is that it is not 
Hermitian and becomes such only in the continuum limit.

As a discretized Hamiltonian is 
generally used to compute the lowest bound states of a system, 
we discuss briefly the error introduced by the discretization in this kind of 
computation.  
In particular we 
focus our attention on the error made when a discretized Hamiltonian acts 
on the eigenfunctions corresponding to the bound states we wish to 
compute. For this purpose we consider the quantity:
\begin{equation}
\Delta_n(x)=\left|\Psi_n^\dagger (x)H_{{\rm CONT}}(x)
\Psi_n (x)-\sum _y \Psi_n^\dagger (x)
H_{{\rm LATT}}(x,y)\Psi_n (y)\right|\quad,
\label{delta}
\end{equation}
where $\Psi_n$ is the exact wave function corresponding to the $n$-th 
eigenvalue of $H_{{\rm CONT}}$ and all quantities are evaluated at a generic 
grid point $x$. 
It is worth noticing that the error $\Delta_n$ has a component due to the 
discretization of the Laplacian operator on the grid and an other due to 
the discretization of the gauge field. The systematic error due to the 
discretization of the spatial derivatives is 
\begin{equation}
\Delta^0_n(x)={1\over 6}\left|
\Psi_n^\dagger (x)\sum_\mu\,{\left. \partial^4\Psi_n\over 
\partial x^4_\mu\right|_x} 
\right|{\hbar^2\over 2 m^*}\,a^2\,. 
\label{er0}
\end{equation}
This error is the same for both Hamiltonians~(\ref{Hlat}) and 
(\ref{hwr}). 
As far as it concerns the error due to the discretization of the gauge field, 
the following considerations apply:
When in Eq.~(\ref{delta}) for $H_{{\rm LATT}}$ we use $H_{{\rm WH}}$
(given by Eq.~\ref{Hlat}), 
both $\Psi_n^\dagger (x)H_{{\rm CONT}}(x)\Psi_n (x)$ and $\sum _y 
\Psi^\dagger (x) H_{{\rm WH}}(x,y)\Psi(y)$ are gauge invariant; 
hence the contribution to $\Delta_n$ due to the vector potential is gauge invariant too and it depends only upon the
coupling between the particle and the field (i.e. $g$), upon the magnetic
field $\vec{B}$ and upon the lattice spacing $a$. 
It may be shown that the component of the error $\Delta_n$ due to the vector potential 
is small 
for any lattice point $x$ if:
\begin{equation}
|ga^2\vec B(x)|\ll 1\quad\quad \forall x\,.
\label{condb}
\end{equation}
Therefore, when  condition~(\ref{condb}) is satisfied and the systematic 
error~(\ref{er0}) is small for the states we wish to compute,  
the Hamiltonian~(\ref{Hlat}) is a good approximation of~(\ref{Hcont}).
We notice that Eq.~(\ref{condb}) is consistent with the fact that the stronger 
the magnetic field the more localized are the eigenfunctions and thus a 
smaller lattice spacing $a$ is required to get a good description of them. 
Instead, if in Eq.~(\ref{condb}) we use $H_{{\rm BDH}}$, then the
quantity $\sum _y \Psi^\dagger (x){H}_{{\rm BDH}}(x,y)\Psi(y)$ 
is no more gauge invariant and as a consequence the error $\Delta_n$ 
depends on the choice of the vector potential $\vec A$. In particular for 
some ``unhappy'' gauge choices the error given by~(\ref{delta}) may be 
so large that~(\ref{hwr}) becomes useless even for a rough estimate 
of the ground state energy. 

As tight-binding Hamiltonians are often used to discretize the Schr\"odinger 
equation 
on a lattice, it is interesting to see how the proposed 
Hamiltonian~(\ref{Hlat}) compares to them. Without a magnetic field, a  
tight binding Hamiltonian (see e.g.~\cite{Sols}) is equivalent 
to that obtained with balanced 
difference methods, although written with another formalism. 
In the presence of a gauge field, a phase shift is introduced in the 
hopping potential~\cite{Mshift}. A tight-binding 
Hamiltonian, for discretizing a Schr\"odinger problem in the presence 
of a magnetic field, reads:
\begin{equation} 
\label{tham}
H_{\rm TB}=\sum_x\,\left[N_d{\hbar^2\over  m a^2}+V(x)\right]|x\rangle\langle 
x|+\sum_x\sum_{y\ne x}\left(-{\hbar^2\over 2 m a^2}\right)e^{i\theta _{x,y}}
|x\rangle\langle y|,
\end{equation}
where $N_d$ is the number of dimensions, the summation over $y$ extends only 
to the next-neighbor of the lattice point $x$, and the phase shift is given by: 
\begin{equation}
\label{pshift}
\theta_{x,y}=-{q\over\hbar}\int_x^y\vec A\cdot \vec { d\mu}, 
\end{equation}
with the integration being performed on the straight line 
connecting the two points $x$ and $y$. 
The first thing to notice is that Hamiltonian~(\ref{tham}) has 
the correct continuum limit and is  
gauge invariant, as can be easily proved. 
The expression~(\ref{pshift}) for the phase shift was 
introduced for the tight-binding computation 
of electronic states in 
crystal lattices. For a crystal, the lattice constant is fixed and depends 
on the particular structure being investigated; therefore the 
gauge field $\vec A$ can vary considerably over a lattice constant and 
thus the use of an integral in the phase shift~(\ref{pshift}) is needed.
This consideration does not apply to the case of a lattice 
of discretization points, as the lattice constant is chosen sufficiently small 
in order to get a good approximation. We also wish to point out that 
the proposed Hamiltonian~(\ref{Hlat}) can be written in the tight-binding 
formalism:
\begin{equation}
\label{htbw}
H_{\rm TBW}=\sum_x\,\left[N_d{\hbar^2\over 2 m a^2}+V(x)\right]|x\rangle\langle
x|+\sum_{x,\mu}\left(-{\hbar^2\over 2 m a^2}\right)\left[
U_{\mu} (x-\mu)|x\rangle\langle x-\mu|+{U_{\mu}}^{\dagger}(x)
|x\rangle\langle x+\mu|\right]\, .
\end{equation}
Equation~(\ref{htbw}) suggests a simpler expression for the hopping potential 
in the presence of a magnetic field. In fact the magnetic field can 
be taken into account by introducing the lattice operator  
$U_\mu$. Moreover, the Hamiltonian~(\ref{htbw}) does not require the 
knowledge of the gauge field over the entire space but just on the lattice 
points; this is very useful if the vector potential is computed 
by some discretization technique. 
Finally, it is worth noticing that, as far as the discretization of 
a Schr\"odinger problem is concerned, 
Hamiltonian~(\ref{htbw}) is practically coincident with~(\ref{tham}), 
although Eq.~(\ref{tham}) is 
more complicated.

\section{Numerical example}
\label{sec2}
In this section, we use Hamiltonians~(\ref{Hlat}) and (\ref{hwr}) to compute 
the lowest part of the energy spectrum and the corresponding eigenfunctions 
for an exactly solvable problem: an electron confined in a two-dimensional 
parabolic potential in GaAs in the presence of a constant perpendicular 
magnetic field of strength $B$. The fact that this problem is analytically 
solvable allows us to test the accuracy of the discretized Hamiltonians and 
their behavior under gauge transformations. 

The Hamiltonian of the system is:
\begin{equation}
\label{Hpar}
H_{{\rm CONT}}={\hbar^2\over 2m^*}
 \, \left[-\nabla ^2  +i g\left\{ \vec A(\vec r\,),\vec{\nabla}  \right\}
+g^2\,\left|\vec A(\vec r\,) \right|^2\right]+{1\over 2}m^*\Omega^2|\vec r-\vec r_C|^2 ,
\end{equation}
where $g=-e/\hbar$ with $e$  modulus of the electron charge, $\hbar\Omega$ is 
the confining energy, $\vec r_C$ is the coordinate of the 
center of the parabolic potential, 
and $m^*$ is the effective mass for an electron in GaAs. 
 
To approach  analytically the problem of computing 
the eigenvalues and the eigenvectors of the 
Hamiltonian~(\ref{Hpar}), it is convenient to use spherical coordinates 
on the plane where the electron is confined with 
the origin at the center of the parabolic potential.

With the gauge choice $\vec A=-(1/2)\vec r\wedge \vec 
B$, the eigenvalues and eigenvectors of~(\ref{Hpar}) are given by~\cite{Fock}:

\begin{eqnarray}
&&E_{n,M}={1\over 2}\hbar \Omega_{eff}(2n+|M|+1)+{1\over 2} \hbar
\omega_c M \label{eigvax}\\
&&\Psi_{n,M}={1\over \lambda\sqrt{2\pi}}\sqrt{{n!\over (n+|M|)!}}
e^{iM\varphi}\exp\left (-{r^2\over 4\lambda^2}\right )\left (
{r^2\over 2\lambda^2}\right )^{|M|\over 2} {L_n}^{|M|}\left(
{r^2\over 2\lambda^2}\right )\, ,
\label{eigvex}
\end{eqnarray}
where $\omega_c=eB/m^*$, 
$\Omega_{eff}^2=4\Omega^2+{\omega_c}^2$, 
$\lambda^2=\hbar/(m^* \Omega_{eff})$, 
$n$ is a non negative integer quantum number, $M$ is the angular 
momentum quantum number and ${L_n}^{|M|}$ is the $n$-th Laguerre polynomial. 

In order to compute numerically the lowest eigenvalues and
eigenvectors of~(\ref{Hpar}) with the discretized Hamiltonians
(\ref{Hlat}) and (\ref{hwr}), we introduce a finite square  grid
centered around the point $\vec r_C$, and we impose Dirichlet boundary 
conditions. This is equivalent to approximating the parabolic potential 
$V$ with a potential 
$\tilde V$ which is equal to $V$ inside the grid region and goes to 
infinity outside it.
We notice that the larger the region covered by the grid the more 
eigenvalues of (\ref{Hpar}) can be effectively approximated. 
In the following part of this section, dealing with the discretized 
computation, we use an orthogonal coordinate system with the grid placed 
in the first quadrant of the $xy$ plane and the $z$-axis perpendicular to it. 
In all the numerical simulations we have taken a confining energy $\hbar\Omega$ 
of 4 meV and we have used for $m^*$ the value $0.067m_0$ with $m_0$ being 
the electron mass. We have used a grid of $99\times 99$ points and a 
lattice constant $a$ of $2.4$ nm, thus covering a region of 
$240$ nm $\times 240$ nm. 

As our purpose is to investigate the behavior of the discretized Hamiltonians 
(\ref{Hlat}) and (\ref{hwr}) under gauge transformations, we compute the 
eigenvalues and the eigenvectors for the following three gauge choices:
\begin{eqnarray}
\mbox{gauge 1:}\quad\quad && A_x=-{y\over 2}\,B\nonumber\\
&& A_y={x\over 2}\,B
\label{gau1} \\
\mbox{gauge 2:}\quad\quad && A_x=-{y-y_C\over 2}\,B\nonumber\\
&&A_y={x-x_C\over 2}\,B
\label{gau2} \\
\mbox{gauge 3:}\quad\quad && A_x=-(x-y)\,B\nonumber\\
&&A_x=y\,B\,.
\label{gau3}
\end{eqnarray}
These gauges are transverse; in particular,  gauge 1 corresponds to 
$\vec{A}=-(1/2)\vec{r}\wedge\vec{B}$, 
while gauge 2 corresponds to $\vec{A}=-(1/2)(\vec{r}-\vec{r}_C)\wedge\vec{B}$. 
Gauge 2 is similar to gauge 1, but symmetric  with respect to the
center of  parabolic potential and, as we shall see below,  it yields more 
accurate results with 
respect  to the other two gauges when inserted in the naively discretized
Hamiltonian~(\ref{hwr}). 
In the following, we shall refer to the Hamiltonians~(\ref{Hlat}) and~
(\ref{hwr}) as WH (Wilson Hamiltonian) and BDH (Balanced Difference 
Hamiltonian) respectively.

In Fig.~1 we show the lowest fifteen eigenenergies for a
magnetic field strength of 10 T, computed with WH
and BDH,
together with the exact ones. We have numbered the eigenvalues with an 
integer $i$  so that $E_i\leq E_{i+1}$ and we plot them versus
$i$. We have found that the results obtained 
using WH with the three 
gauges~(\ref{gau1},\ref{gau2},\ref{gau3}) do not differ; therefore for the
quantities computed with WH we do not specify which gauge 
we have used. From the above mentioned figure, we notice  
that: (1) The eigenvalues computed with BDH using
gauges 1 and 3 differ appreciably from the exact ones; 
in particular, this difference grows with
increasing  eigenvalue. (2) The eigenvalues computed with BDH using
gauge 2 tend to differ from the the exact ones for the excited
states. 

The explicit breaking of gauge invariance present in the Hamiltonian
can be further investigated analyzing the probability density
$|\Psi_{n,M}|^2$: in Fig.~2 we show the probability density 
for the ground state (with $B=10$ T) using either the corresponding exact
eigenfunction (from Eq.~\ref{eigvex}) or that obtained with BDH; 
for this computation 
we have not found significant differences 
between the exact probability density and either that computed 
with WH (using the three different gauges) or that computed 
with BDH using gauge 2. From Fig~2 it is evident 
that in this case the breaking of gauge-invariance of BDH
``disrupts'' the symmetry of the probability density with respect to the
center of the confining potential. We have found the same phenomenon
with the gauge 3, while this does not occur  
using gauge 2 because this gauge preserves the above mentioned symmetry. 

In Fig.~3 we report the probability density for the 14-th
excited state, i.e. that with  $n=0,M=-14$ (see Eq.~\ref{eigvax}), and $B=10$T 
using the corresponding exact eigenfunction and that computed
with BDH using gauge 2; also in this case we have not found
significant differences between the exact calculation and the one
performed with WH, while it is evident that
the result obtained with BDH using gauge 2 does not match the
exact one. At this point it is worth stressing that even if in some cases, 
for a careful gauge choice, BDH gives 
a good estimate of the eigenenergies, the estimate of the corresponding wave 
functions may be inadequate (see Fig~1 and Fig.~3). 
This implies that the matrix elements of an operator corresponding to an 
observable physical quantity (e.g. dipole and quadrupole momenta) between the 
eigenfunctions computed with BDH can be affected by large errors.

Finally, in Fig.~4 we show the ground state
energy as a function of the magnetic field magnitude computed with the 
exact expression and with the three different gauges using both 
WH and BDH. The results obtained 
using WH with the three 
gauges~(\ref{gau1},\ref{gau2},\ref{gau3}) do not differ; therefore for the
quantities computed with WH the gauge used is not specified. 
This analysis shows how the effect of the breaking of the gauge
invariance in the spectrum of BDH grows with increasing
magnetic field magnitude. 

\section{Conclusions}
\label{sec3}
In this paper we have addressed the problem of developing a method to find 
a gauge invariant grid discretization of a one-particle Hamiltonian in the presence
of an external electromagnetic field; for this purpose, we have used
the Wilson formulation of lattice gauge theories. 
This Hamiltonian, written in the tight-binding formalism, 
leads to a simpler expression for the hopping potential, that does not 
require the knowledge of the gauge field over the entire space but only 
on the lattice of discretization points.
The method developed
has been checked in an exact solvable case, i.e. a Hamiltonian
describing an electron confined in a two-dimensional 
parabolic potential in the presence of a constant perpendicular 
magnetic field. The results obtained clearly show how a ``na\"{\i}ve''
discretization yields to an explicit breaking of the gauge invariance
and may lead to large errors in the estimates of the eigenvalues and
corresponding probability densities. 
With a careful gauge choice a naively discretized Hamiltonian can be used 
to give a rough estimate of the energy spectrum, 
but the corresponding eigenfunctions are still affected by large errors, 
thus being unsuitable for the computation of the matrix elements of
operators corresponding to relevant physical quantities.
On the other hand, the proposed
discretized Hamiltonian~(\ref{Hlat}) has the correct behavior 
under gauge transformations and allows an accurate computation of
both the energy spectrum and the probability densities. 

\section*{ACKNOWLEDGMENTS}

We thank Prof.~Pietro Menotti for his very useful suggestions and
illuminating discussions. 

This work has been supported by the ESPRIT Project 23362 QUADRANT
(QUAntum Devices foR Advanced Nano-electronic Technology).

\newpage
\begin{figure}
\protect\caption{Lowest fifteen eigenvalues for $B=10$T computed with: exact
Hamiltonian (points); WH (diamonds); BDH with gauge 1 (circles); BDH
with gauge 2 (cross);  BDH with gauge 3 (squares).}
\end{figure}
\begin{figure}
\protect\caption{Ground state probability density for $B=10$T: a) exact; b)
computed with BDH (gauge 1).}
\end{figure}
\begin{figure}
\protect\caption{Probability density of the $14$-th excited state ($n=0$,
$M=-14$): a) exact; b) computed with BDH (gauge 2).}
\end{figure}
\begin{figure}
\protect\caption{Ground state energy plotted against magnetic field magnitude 
computed with: 
exact Hamiltonian (continuous line); WH (diamonds); 
BDH with gauge 1 (circles); BDH with gauge 2 (cross);  
BDH with gauge 3 (squares).}
\end{figure}
\end{document}